\documentclass[10pt]{elsart}
\usepackage[T1]{fontenc}
\usepackage{graphics}

\begin{document}

\begin{frontmatter}
\title{Angular momentum projected analysis of Quadrupole Collectivity in \protect\( ^{30,32,34}Mg\protect \)
and \protect\( ^{32,34,36,38}Si\protect \) with the Gogny interaction.}

\author{R. Rodr\'\i guez-Guzm\'an, J.L. Egido and L.M. Robledo.}

\address{Departamento de F\'\i sica Te\'orica C--XI, Universidad Aut\'onoma de Madrid, 
28049 Madrid, Spain }

\maketitle
\begin{abstract}
A microscopic angular momentum projection after variation is used to describe
quadrupole collectivity in \( ^{30,32,34}Mg \) and \( ^{32,34,36,38}Si \).
The Hartree-Fock-Bogoliubov states obtained in the quadrupole constrained mean
field approach are taken as intrinsic states for the projection. Excitation
energies of the first \( 2^{+} \) states and the \( B(E2,0^{+}\rightarrow 2^{+}) \)
transition probabilities are given. A reasonable agreement with available experimental
data is obtained. It is also shown that the mean field picture of those nuclei
is strongly modified by the projection. 
\end{abstract}
\end{frontmatter}

Neutron-rich nuclei with \( N\approx 20 \) are spectacular examples of shape
coexistence between spherical and deformed states. Experimental evidence for
an island of deformed nuclei near \( N=20 \) has been found in the fact that
\( ^{31}Na \) and \( ^{32}Na \) are more tightly bound than could be explained
with spherical shapes \cite{ray.1}. Additional support comes from the unusually
low excitation energy of the \( 2^{+} \) state in \( ^{32}Mg \) \cite{exp32mg}.
A large ground state deformation has also been inferred from intermediate energy
Coulomb excitation studies \cite{ray.2} in \( ^{32}Mg \). Quadrupole collectivity
of \( ^{32-38}Si \) has also been studied in \cite{ray.3}. Very recently this
region has been the subject of detailed experimental spectroscopic studies at
ISOLDE \cite{ray.4}. From a theoretical point of view, deformed ground states
have been predicted for nuclei with \( N\approx 20 \) \cite{ray.5,ray.6,ray.7}.
In those calculations the rotational energy correction is the essential ingredient
for the stabilization of the deformed configuration. On the other hand, some
calculations have predicted \cite{ray.8,ray.9,ray.10,ray.11} a spherical ground
state for \( ^{32}Mg \) but it has also been found \cite{ray.11} that deformation
effects may appear as a result of dynamical correlations. Some shell model calculations,
even with restricted configuration spaces, have been able to explain the increased
quadrupole collectivity at \( N=20 \) as a result of neutron \( 2p-2h \) excitations
into the \( fp \) shell, see for example \cite{ray.12,ray.13}. Recently, a mean
field study has explored the suitability of several Skyrme parameterizations
\cite{Rein.99} in the description of this and other regions of shape coexistence.

The mean field description of nuclei is usually a good starting point as it
provides a qualitative, and in many cases quantitative, understanding of the
nuclear properties. This is the case when the mean field solution corresponds
to a well defined minimum. However, in regions of shape coexistence where two
minima are found at a comparable energy, the correlation effects stemming from
the restoration of broken symmetries and/or collective motion can dramatically
alter the energy landscape thus changing the mean field picture. For this reason,
we have included in our mean field calculations the effects related to the restoration
of the broken rotational symmetry by performing, for the first time with the
Gogny force, angular momentum projected calculations of the energies and other
relevant quantities. The reason for choosing to restore rotational symmetry
is that the zero point energy associated with this restoration is somehow proportional
to deformation and ranges, in this region, from a few KeV for nearly spherical
configurations to several MeV for well deformed ones. This energy range
is comparable to the energy differences found between different shapes in nuclei
of this region. Therefore, in addition to the mean field results, both angular momentum
projected \( I=0 \) and \( I=2 \) surfaces were computed for the nuclei \( ^{30,32,34}Mg \)
and \( ^{32,34,36,38}Si \) and angular momentum projected transition
probabilities \( B(E2,0^{+}\rightarrow 2^{+}) \) among different configurations.

The calculation proceeds in two steps: in the first one we perform a set of
constrained Hartree- Fock- Bogoliubov (HFB) calculations using the D1S parameterization
\cite{Berg.84} of the Gogny force \cite{ray.14} and the mass quadrupole operator
\( \hat{Q}_{20}=z^{2}-\frac{1}{2}(x^{2}+y^{2}) \) as the constraining operator
in order to obtain a set of ``intrinsic'' wave functions \( \mid \phi (q_{20})\rangle  \).
The self-consistent symmetries imposed in the calculation were axial symmetry,
parity and time reversal. The two body kinetic energy correction was fully taken
into account in the minimization process. On the other hand, the Coulomb exchange
term was replaced by the local Slater approximation and neglected in the variational
process. The Coulomb pairing term as well as the contribution to the pairing
field from the spin-orbit interaction were neglected. A harmonic oscillator (HO)
basis of 10 major shells was used to expand the quasi-particle operators and
the two oscillator lengths defining the axially symmetric HO basis were kept
equal for all the values of the quadrupole moment. The reason for choosing the
basis this way was that we wanted a basis closed under rotations (i.e.
an arbitrary rotation of the basis elements always yields wave functions that
can be solely expressed as linear combinations of the elements of the basis)
in order to avoid the technical difficulties discussed in \cite{ray.15} when
a non-closed basis is used. In the second step we compute the angular momentum
projected energy for each intrinsic wave function \( \mid \phi (q_{20})\rangle  \)
obtaining in this way a set of energy curves \( E_{I}(q_{20}) \) for each value
of \( I=0,2,\ldots  \) The minima of each curve provide us with the energies
and wave functions of the \( I=0^{+},2^{+},\ldots  \) yrast and isomeric states.

The theoretical background for angular momentum projection is very well described
in \cite{ray.18,Hara.95} and therefore we will not dwell on the details here.
However, a few remarks concerning the peculiarities of our calculation are in
order: first, and due to the axial symmetry imposed in the HFB wave functions,
the angular momentum projected energy is given by

\begin{equation}
\label{EPROJ}
E_{I}(q_{20})=\frac{\int _{0}^{\frac{\pi }{2}}d\beta sen\beta d_{00}^{I}(\beta )\langle \phi (q_{20})\mid \hat{H}'[\rho _{\beta }(\vec{r})]e^{-i\beta \hat{J}_{y}}\mid \phi (q_{20})\rangle }{\int _{0}^{\frac{\pi }{2}}d\beta sen\beta d_{00}^{I}(\beta )\langle \phi (q_{20})\mid e^{-i\beta \hat{J}_{y}}\mid \phi (q_{20})\rangle }
\end{equation}
 with \( \hat{H}'[\rho _{\beta }(\vec{r})]=\hat{H}[\rho _{\beta }(\vec{r})]-\lambda _{\pi }(\hat{N}_{\pi }-Z)-\lambda _{\nu }(\hat{N}_{\nu }-N) \).
The term \( -\lambda _{\pi }(\hat{N}_{\pi }-Z)-\lambda _{\nu }(\hat{N}_{\nu }-N) \)
is included to account for the fact that the projected wave function does not
have the right number of particles on the average. The previous term would correspond
to the application of first order perturbation theory if the chemical potentials
used were the derivatives of the projected energy with respect to the number
of particles. In our calculations we have simply used the chemical potentials
obtained in the HFB theory\footnote{%
This recipe has been previously used \cite{ray.19} in the context of angular
momentum projection and in Generator Coordinate Method (GCM) calculations 
\cite{Bon.90}. In both cases it has been found that the present recipe works 
very well.}. This simplification is justified by the fact that the deviations induced in
the number of particles due to the angular momentum projection are always small
and so are their effects on the projected energies. 

For the computation of the matrix elements of the rotation operator in a HO
basis we have used the results of ref. \cite{ray.16}. 

Another relevant point to be discussed is the prescription to use for the density
dependent part of the Gogny force. In the calculation of the energy functional
\( E[\phi ]=\left\langle \phi \right| \hat{H}\left| \phi \right\rangle  \)
the density appearing in the density dependent part of the force is simply \( \rho (\vec{r})=\left\langle \phi \right| \hat{\rho }\left| \phi \right\rangle  \)
rendering the energy a functional of the density and the pairing tensor but
with a dependence on the density different from the canonical quadratic one
of the standard HFB theory. On the other hand, the energy overlap \( E[\phi ,\phi ']=\left\langle \phi \right| \hat{H}\left| \phi '\right\rangle /\left\langle \phi |\phi '\right\rangle  \)
can be evaluated using the extended Wick theorem. The final expression is the
same as the HFB functional \( E[\phi ] \) but replacing the density matrix
by \( \bar{\rho }_{ij}=\left\langle \phi \right| c_{j}^{+}c_{i}\left| \phi '\right\rangle /\left\langle \phi |\phi '\right\rangle  \)
and the pairing tensor by \( \bar{\kappa }_{ij}=\left\langle \phi \right| c_{j}c_{i}\left| \phi '\right\rangle /\left\langle \phi |\phi '\right\rangle  \)
and \( \overline{\tilde{\kappa }}_{ij}=\left\langle \phi \right| c_{i}c_{j}
\left| \phi '\right\rangle /\left\langle \phi |\phi '\right\rangle  \). As a consequence, it seems rather natural to use the density \( \bar{\rho }\left( \vec{r}\right) =\left\langle \phi \right| \hat{\rho }\left| \phi '\right\rangle /\left\langle \phi |\phi '\right\rangle  \)
in the evaluation of the density dependent term of the force in \( E[\phi ,\phi '] \).
In our case, this leads to the introduction of a density dependent term depending
on \( \bar{\rho }\left( \vec{r},\beta \right) =\left\langle \phi \right| \hat{\rho }e^{-i\beta \hat{J}_{y}}\left| \phi \right\rangle /\left\langle \phi \right| e^{-i\beta \hat{J}_{y}}\left| \phi \right\rangle  \).
This density dependence seems to yield to bizarre consequences like having a
non-hermitian and non rotationally invariant hamiltonian. These apparent inconsistencies
can be overcome if we think of a density dependent force, not as an operator
to be added to the kinetic energy in order to obtain a hamiltonian, but rather
as a device to get energy functionals like \( E[\phi ] \) and \( E[\phi ,\phi '] \)
with the property of yielding an energy that is a real quantity and independent
of the orientation of the reference frame. The density dependence just mentioned
fulfills these two requirements as can be readily checked. In addition, when
the intrinsic wave function is strongly deformed and the Kamlah expansion can
be used to obtain an approximate expression for the projected energy (the cranking
model) the above density dependence yields the correct expression for the angular
velocity \( \omega  \) including the ``rearrangement'' term \cite{Val.82}.
A more elaborated argumentation in favor of the density dependence just mentioned
will be given elsewhere.
\begin{figure}
\par\centering \resizebox*{10cm}{!}{\includegraphics{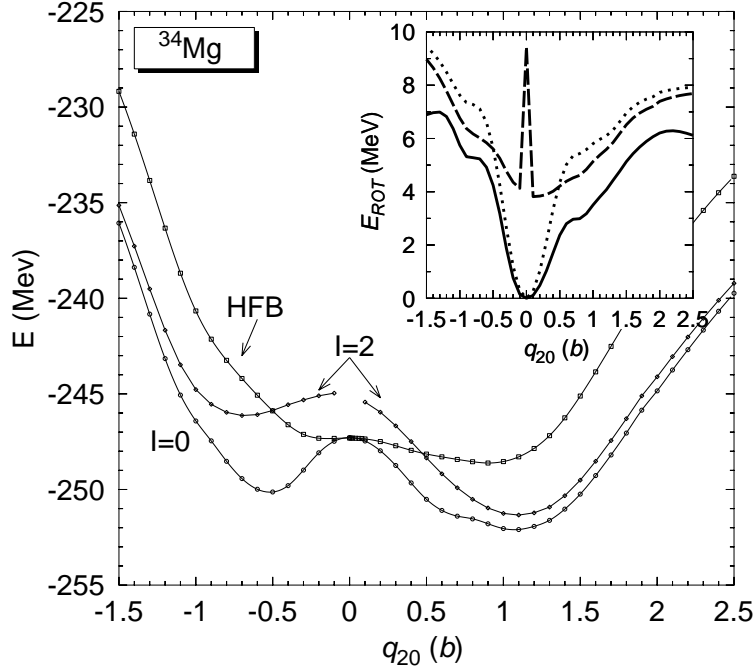} } \par{}

\caption{The HFB and projected energies as a function of \protect\protect\( q_{20}\protect \protect \)
for the nucleus \protect\protect\( ^{34}Mg\protect \protect \). See text for
further details.\label{figure.1}}
\end{figure}

As an example of the results obtained we show in Fig. \ref{figure.1} the HFB
and projected energies as a function of \( q_{20} \) for the nucleus \( ^{34}Mg \).
In contrast with the HFB result, the \( I=0 \) energy surface shows two pronounced
minima in the prolate and oblate side which are rather close to each other in
energy being the prolate minimum slightly deeper than the oblate one. Therefore,
it is difficult to assign a given character to the \( I=0 \) state until a
configuration mixing calculation is performed, although it is very likely that
the predominant configuration for the \( I=0 \) state is going to be the prolate
one. For \( I=2 \) there is a well developed prolate minimum. Let us also mention
that for configurations with a \( q_{20} \) value close to zero (i.e. close
to the spherical configuration \( q_{20}=0 \) which is a pure \( I=0 \) state)
it is very difficult to compute the \( I=2 \) projected energy due to numerical
instabilities related to the smallness of \( \left\langle \phi (q_{20})\right| \hat{P}_{00}^{I}\left| \phi (q_{20})\right\rangle  \). 

In the inset of Fig. \ref{figure.1} we have plotted the energy difference \( E_{ROT}(I)=E_{HFB}-E_{I} \)
as a function of \( q_{20} \) for \( I=0 \) (full line) in order to compare
it with the rotational energy correction \( E^{App}_{ROT}=\left\langle J^{2}_{y}\right\rangle /\mathcal{J}_{Y} \)
often used in mean field calculations (dashed line). The Yoccoz moment of inertia
\( \mathcal{J}_{Y} \) has been computed, as it is usually done, in an approximate
way by neglecting the two body quasiparticle interaction term of the hamiltonian
(the same kind of approximation yields to the Inglis-Belyaev moment of inertia
instead of the Thouless-Valatin one). We notice that \( E^{App}_{ROT} \) agrees
qualitatively well with \( E_{ROT}(0) \) for \( q_{20} \) values greater than
\( 100fm^{2} \) and smaller than \( -50fm^{2} \) as expected: these are regions
of strong deformation where the validity conditions for \( E^{App}_{ROT} \)
to be a good approximation to \( E_{ROT} \) (Kamlah expansion) are satisfied.
On the other hand, the behavior of \( E^{App}_{ROT} \) is completely wrong
in the inner region. One prescription to extend the rotational formula to weakly
deformed states is the one of \cite{Rein.87} based on results with the Nilsson
model. The prescription multiplies \( E_{ROT}^{App} \) by a function of \( \left\langle J_{y}^{2}\right\rangle  \)
with the property of going to zero (one) for \( \left\langle J_{y}^{2}\right\rangle  \)
going to zero (infinity). The resulting rotational energy is also depicted in
the inset of Fig. \ref{figure.1} (dotted line) and, although the qualitative
agreement with the exact result improves somewhat, the quantitative one is far
from satisfactory in the nucleus considered. The rotational energy correction
formula is based on the assumption that the quantity \( h(\beta )=\left\langle \hat{H}e^{-i\beta \hat{J}_{y}}\right\rangle /\left\langle e^{-i\beta \hat{J}_{y}}\right\rangle  \)
can be very well approximated by a quadratic function \( h(\beta )\approx h(0)+\frac{1}{2}h''(0)\beta ^{2} \)
where \( h''(0) \) is related to the exact Yoccoz moment of inertia by the
expression \( \mathcal{J}_{Y}=-\left\langle \hat{J}^{2}_{y}\right\rangle ^{2}/h''(0) \).
It is well known that this assumption is justified for deformed heavy nuclei.
However, we have checked that it is not the case for the nuclei studied here
even for the largest deformations considered. Therefore, we conclude that the
exact restoration of the rotational symmetry is fundamental for a qualitative
and quantitative description of the rotational energies in these light nuclei.
\begin{figure}
\resizebox*{14cm}{7cm}{\includegraphics{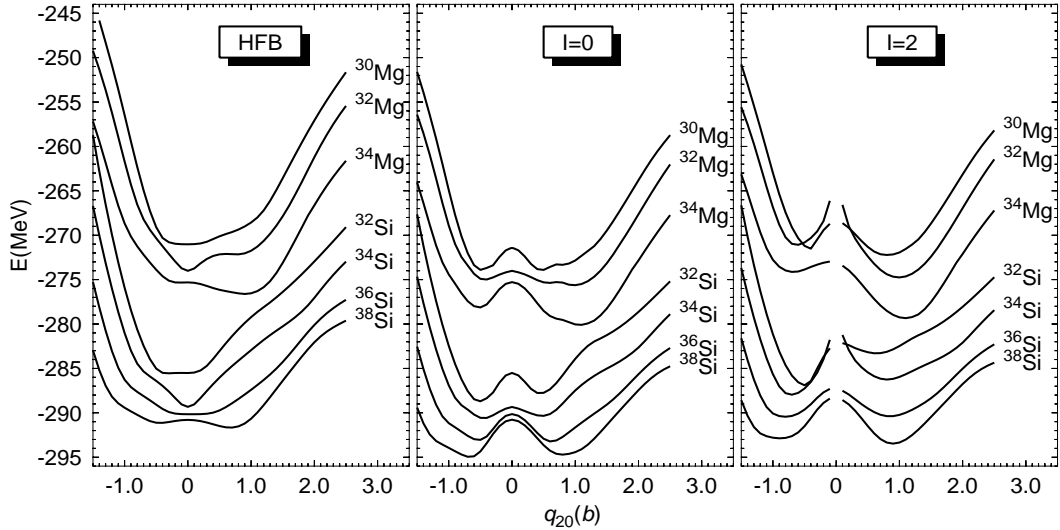} }
\caption{HFB (left), angular momentum projected \protect\( I=0\protect \) (middle)
and \protect\( I=2\protect \) (right) collective potential energy surfaces
for the nuclei \protect\protect\( ^{30,32,34}Mg\protect \protect \) and \protect\protect\( ^{32,34,36,38}Si\protect \protect \).
All the curves have been shifted by the same amount for each nucleus in all
the three cases in order to fit them in a single plot. The energy shifts are
-36, -30, and -28 MeV for \protect\protect\( ^{30,32,34}Mg\protect \protect \)
and -23, -13 and -6 MeV for \protect\protect\( ^{32,34,36}Si\protect \protect \)
respectively. The range of quadrupole moments considered roughly correspond
to a \protect\protect\( \beta _{2}\protect \protect \) range from -0.5 to 0.85. }
\label{figura2}
\end{figure}

The main outcomes of the calculation are summarized in Fig. \ref{figura2} where
we show, on the left hand side panel, the HFB potential energy surfaces for
Mg and Si isotopes as a function of the mass quadrupole moment. These surfaces
have been shifted accordingly to fit them in the plot. We observe that only
in the nuclei \( ^{34}Mg \) and \( ^{38}Si \) we obtain a prolate minimum
at \( \beta _{2} \) deformations of 0.4 and 0.35 respectively. For the other
nuclei, the minimum corresponds to the spherical configuration. For all the
nuclei considered the energy curves are very flat around the corresponding minimum
indicating that further correlations can substantially modify the energy landscape
and therefore the conclusions obtained from the raw HFB results.

On the middle and right hand side panels of Fig. \ref{figura2} we show the
angular momentum projected \( I=0 \) and \( I=2 \)  potential energy
surfaces for all the nuclei considered. These surfaces have also been shifted
to fit them in the plot. For \( I=0 \), apart from the nucleus \( ^{34}Mg \)
that shows a rather clear prolate minimum, the general trend for the ground
state is to show shape coexistence. For \( I=2 \) we have prolate minima for
\( ^{32-34}Mg \) and oblate minima for \( ^{32-34}Si \) whereas the other
nuclei are examples of shape coexistent structures. The results just shown indicate
that, for a quantitative description of the ground and \( 2^{+} \) states in
all these nuclei, a configuration mixing calculation (GCM) using the mass quadrupole
moment as generating coordinate is needed. In spite of this, we present in Table
\ref{tabla2} the \( 0^{+}-2^{+} \) energy differences for the four possible
configurations with the \( 0^{+} \) in the prolate (P) or oblate (O) minimum
and the \( 2^{+} \) also in the P or O minimum. The energies written in boldface
correspond to the predictions obtained by strictly using the criterion of the
absolute minimum of \( E_{I}(q_{20}) \) to assign the \( 0^{+} \) and \( 2^{+} \)
states. Comparison of these predictions with the experimental results indicates
a reasonable agreement except for \( ^{36}Si \). The inclusion of configuration
mixing will presumably improve the agreement as it always yields to a mixed
configuration with an energy lower than the energies of the states being mixed.
Therefore, if the \( 0^{+} \)configurations strongly mix (shape coexistence)
but the \( 2^{+} \)ones do not ( there is a well established minima) the \( 0^{+}-2^{+} \)
energy difference will increase whereas it will decrease if the opposite situation
takes place. On the other hand, if configuration mixing is important for both
the \( 0^{+} \) and \( 2^{+} \) states anything can happen to the excitation
energy. Therefore, we expect that configuration mixing is going to increase
the excitation energies in all cases except in \( ^{30}Mg \) and \( ^{36,38}Si \)
where the behavior is unpredictable. 
\begin{table}

\caption{Excitation energies, in MeV, for the \protect\protect\( 2^{+}\protect \protect \)
states. The four possible combinations are shown. The numbers in boldface indicate
the configuration where the \protect\protect\( 0^{+}\protect \protect \) and
the \protect\protect\( 2^{+}\protect \protect \) correspond to the absolute
minimum of the projected energy surface.The experimental data are taken from
\cite{ray.3} (Si) and from \cite{exp32mg} (Mg).}
\begin{tabular}{rccccc}
\hline 
&
 \( (0^{+}_{P}-2^{+}_{P}) \)&
 \( (0^{+}_{O}-2^{+}_{P}) \)&
 \( (0^{+}_{P}-2^{+}_{O}) \)&
 \( (0^{+}_{O}-2^{+}_{O}) \)&
 Exp\\
\hline 
\( ^{30}Mg \)&
 \textbf{1.683}&
 1.681 &
2.391 &
 2.388 &
\\
 \( ^{32}Mg \)&
\textbf{0.873}&
 0.235 &
 4.546 &
 3.909 &
 0.885\\
 \( ^{34}Mg \)&
 \textbf{0.753}&
  -1.206 &
  5.960 &
4.000 &
\\
\hline 
\( ^{32}Si \)&
 4.475 &
 5.408 &
 0.869 &
 \textbf{1.803}&
 1.941\\
 \( ^{34}Si \)&
 4.088 &
4.324 &
 2.383 &
 \textbf{2.619}&
 3.327\\
 \( ^{36}Si \)&
  2.845 &
2.670 &
  \textbf{2.778} &
2.603 &
 1.399\\
 \( ^{38}Si \)&
 1.238 &
 \textbf{1.481}&
  1.837 &
 2.080 &
 1.084 \\
\hline 
\end{tabular}
\label{tabla2}
\end{table}
 In Table \ref{tabla3} we present the results obtained for the \( B(E2,0^{+}\rightarrow 2^{+}) \)
transition probabilities for the four possible combinations. As in the previous
table, the results obtained by choosing for the \( 0^{+} \) and \( 2^{+} \)
states the ones corresponding to the absolute minima of the projected energies
are written in boldface. For the nuclei \( ^{32}Mg \) and \( ^{34}Mg \) we
obtain very collective values for the \( B(E2) \) which, in the case 
of \( ^{32}Mg \),
are in rather good agreement with the experiment. For both nuclei, we expect
a contamination of the ground state wave function by the oblate \( 0^{+} \)
state that will yield to a reduction of the \( B(E2) \) values (see column
two for the \( B(E2,0_{O}^{+}\rightarrow 2_{P}^{+}) \)) that will bring the
theoretical predictions in closer agreement with the experimental data. For the
\( ^{32}Si \), \( ^{34}Si \) and \( ^{38}Si \) isotopes we underestimate
the \( B(E2) \) values but, presumably, admixtures of the \( 0^{+}_{P}\rightarrow 2^{+}_{P} \)
transition will help to bring the theoretical results in closer agreement with
the experiment, specially for the \( ^{38}Si \) nucleus. Concerning \( ^{36}Si \)
we can only conclude that a strong \( 0^{+}_{P}\rightarrow 2^{+}_{P} \) component
has to be present in the evaluation of the \( B(E2). \)
\begin{table}

\caption{Transition probabilities \protect\protect\( B(E2,0^{+}\rightarrow 2^{+})\protect \protect \)
in \protect\protect\( e^{2}fm^{4}\protect \protect \) between different configurations
in \protect\protect\( ^{30,32,34}Mg\protect \protect \) and \protect\protect\( ^{32,34,36,38}Si\protect \protect \).
As in the previous table \protect\protect\( O\protect \protect \) (\protect\protect\( P\protect \protect \))
stands for the oblate (prolate) configuration. The experimental data is taken
from \cite{ray.3} (Si) and from \cite{ray.2} (Mg).}
\begin{tabular}{rccccc}
\hline 
&
 \( 0^{+}_{P}\rightarrow 2^{+}_{P} \)&
 \( 0^{+}_{O}\rightarrow 2^{+}_{P} \)&
 \( 0^{+}_{P}\rightarrow 2^{+}_{O} \)&
 \( 0^{+}_{O}\rightarrow 2^{+}_{O} \)&
 Exp.\\
\hline 
\( ^{30}Mg \)&
 \textbf{182.11}&
 39.44&
 0.88&
 8.38 &
\\
 \( ^{32}Mg \)&
 \textbf{593.24}&
 15.53&
 2.26 &
 6.52 &
 454\( \pm  \)78\\
 \( ^{34}Mg \)&
 \textbf{549.21}&
 14.74 &
 2.66 &
 10.51 &
\\
\hline 
\( ^{32}Si \)&
 402.09 &
 54.19 &
 4.26 &
 \textbf{50.71}&
 113\( \pm  \)33\\
 \( ^{34}Si \)&
 227.18 &
 62.51 &
 6.81 &
 \textbf{39.16}&
 85\( \pm  \)33\\
 \( ^{36}Si \)&
 232.22 &
 45.05 &
 \textbf{6.43} &
 28.31 &
 193\( \pm  \)59\\
 \( ^{38}Si \)&
 418.91 &
 \textbf{35.18}&
 4.43 &
 69.80 &
 193\( \pm  \)71 \\
\hline 
\end{tabular}
\label{tabla3}
\end{table}

In Table \ref{tabla1} the HFB and projected ground state energies for the
nuclei under consideration are shown and compared to the experimental data taken
from \cite{ray.17}. The inclusion of the zero point energy stemming from the
restoration of the rotational symmetry clearly improves the theoretical description
of the binding energies.
\begin{table}

\caption{Ground state energies in MeV as compared to the experimental results. The quantities
\protect\protect\( \delta E_{I=0}=E_{Exp}-E_{I=0}-E_{C.E.}\protect \protect \)
and \protect\protect\( \delta E_{HFB}=E_{Exp}-E_{HFB}-E_{C.E.}\protect \protect \)
are also presented. The quantity \protect\protect\( E_{C.E.}\protect \protect \)
stands for the HFB Coulomb exchange energy computed in the Slater approximation.
The experimental data are taken from \cite{ray.17}.}
\begin{tabular}{rcccccc}
\hline 
&
 \( E_{HFB} \)&
 \( E_{I=0} \)&
 \( E_{C.E.} \)&
 \( E_{Exp} \)&
 \( \delta E_{I=0} \)&
 \( \delta E_{HFB} \)\\
\hline 
\( ^{30}Mg \)&
 -235.01 &
 -237.90 &
 -4.29 &
 -241.63 &
 0.56 &
 -2.33\\
 \( ^{32}Mg \)&
 -244.00 &
 -245.62 &
 -4.22 &
 -249.68 &
 0.16 &
 -1.46\\
 \( ^{34}Mg \)&
 -248.61 &
 -252.09 &
 -4.21 &
 -256.58 &
 -0.28 &
 -3.76\\
\hline 
\( ^{32}Si \)&
 -262.55 &
 -265.70 &
 -5.13 &
 -271.41 &
 -0.58 &
 -3.73\\
 \( ^{34}Si \)&
 -276.32 &
 -277.59 &
 -5.06 &
 -283.42 &
 -0.77 &
 -2.04\\
 \( ^{36}Si \)&
 -284.15 &
 -287.22 &
 -5.04 &
 -292.01 &
0.25 &
 -2.82\\
 \( ^{38}Si \)&
 -291.66 &
 -294.95 &
 -4.98 &
 -299.50 &
0.43 &
 -2.86  \\
\hline 
\end{tabular}
\label{tabla1}
\end{table}

In conclusion, we have computed several properties of neutron rich \( Mg \)
and \( Si \) isotopes using the HFB theory and exact angular momentum projection.
In the calculations the finite range density dependent Gogny force has been
used. The results for the excitation energies \( 0^{+}-2^{+} \) and \( B(E2,0^{+}\rightarrow 2^{+}) \)
transition probabilities obtained from the angular momentum projected wave functions
are in reasonable agreement with the experiment. The analysis of the projected
energy surfaces and also the discrepancies found between theory and experiment
indicate that configuration mixing is an important ingredient in these nuclei.
Work is in progress in order to incorporate such configuration mixing.

One of us (R. R.-G.) kindly acknowledges the financial support received from
the Spanish Instituto de Cooperacion Iberoamericana (ICI). This work has been
supported in part by the DGICyT (Spain) under project PB97/0023.

\end{document}